\documentclass{ws-ijmpb}
\begin{document}
\markboth{Gao {\it et al}.}
{Effect of disorder on the interacting Fermi gases in a one-dimensional
optical lattice}

%

\catchline{}{}{}{}{}

%


\title{EFFECT OF DISORDER ON THE INTERACTING FERMI GASES IN A 
ONE-DIMENSIONAL OPTICAL LATTICE}

\author{GAO XIANLONG}
\address{Department of Physics, Zhejiang Normal University, Jinhua, 
Zhejiang Province, 321004, China} 
\author{M. POLINI and M. P. TOSI}
\address{NEST-CNR-INFM and Scuola Normale Superiore, I-56126 Pisa, Italy}
\author{B. TANATAR$^*$}
\address{Department of Physics, Bilkent University, Ankara, 06800, Turkey\\
$^*$E-mail: tanatar@fen.bilkent.edu.tr}

\maketitle

\begin{history}
\received{Day Month Year}
\revised{Day Month Year}
\end{history}

\begin{abstract}
Interacting two-component Fermi gases loaded in a one-dimensional 
(1D) lattice and subjected to an harmonic trapping potential exhibit 
interesting compound phases in which fluid regions coexist with local 
Mott-insulator and/or band-insulator regions. 
Motivated by experiments on cold atoms inside disordered optical 
lattices, we present a theoretical study of the effects of a 
correlated random potential on these ground-state phases. 
We employ a lattice version of density-functional theory 
within the local-density approximation to determine the density 
distribution of fermions in these phases. 
The exchange-correlation potential is obtained from
the Lieb-Wu exact solution of Fermi-Hubbard model. On-site disorder
(with and without Gaussian correlations) and harmonic trap are 
treated as external potentials.
We find that disorder has 
two main effects: (i) it destroys the local insulating regions 
if it is sufficiently strong compared with the on-site atom-atom 
repulsion, and (ii) it induces an anomaly in the inverse compressibility
at low density from quenching of percolation. For sufficiently
large disorder correlation length the enhancement in the inverse
compressibility diminishes.
\end{abstract}


\section{Introduction}

Disorder and interaction effects in condensed matter
systems have a long and rich history. The interplay between 
them has been the subject of continuing interest.
The notable examples range from two-dimensional electron
systems with long-ranged Coulomb interactions\cite{MIT} to liquid
$^4$He absorbed in various substances such as aerogel and
vycor\cite{helium_four} and granular superconductors.\cite{fazio_2001}
In the former system experimental and theoretical
investigations\cite{spin_susceptibility} reveal the crucial 
role played by disorder as the two-dimensional electron system 
undergoes a metal-insulator transition. Furthermore, thermodynamic
quantities carry the signature of phase 
transition\cite{compressibility_anomaly,tanatar_ijmpb} along with
transport properties.

In recent years, cold atomic systems are being used to
investigate the interplay between single-particle randomness 
and interaction effects with the help of optical lattices
allowing access to strong coupling regimes through the
depression of kinetic energy.\cite{speckle,optical_lattices,clement}
Furthermore, the on-site interaction can be tuned either indirectly
by changing the strength of the lasers that create the optical lattice
or directly by means of a Feshbach resonance.\cite{moritz_2005}

After the initial observation of superfluid to Mott
insulator transition using bosonic atoms in an optical
lattice\cite{greiner} and creation of disorder potentials,\cite{hannover}
studies on fermionic atoms\cite{moritz_2005,fermi_lattice}
are beginning to be explored. A couple of forthcoming reviews\cite{giorgini}
encompass many aspects of the theory of ultracold Fermi gases.

In this work we study the interaction and disorder
effects on a one-dimensional, two-component Fermi gas trapped 
in a harmonic confinement potential and an optical lattice. 
We use the lattice version of density functional theory\cite{soft}
taking advantage of the exact solution of the one-dimensional
Fermi-Hubbard model.\cite{lieb_wu} Ground-state calculations
in the absence of disorder have been performed within
a number of numerical techniques\cite{repulsive,gao_long_2005} to 
identify various phases.
In the present approach, the harmonic confinement 
potential and the disorder potential are treated as part of the
Kohn-Sham potential within a local-density approximation
which has an exact representation for the uniform case.

Generalizing our earlier work\cite{gao_short} on the effects of 
uncorrelated disorder, we here study correlated disorder 
which depends on the correlation length as a parameter.
The site occupation profiles and thermodynamic stiffness are 
determined to study the effects of disorder. Our motivation for
considering correlated disorder comes from the experiments\cite{speckle}
on Bose-Einstein condensates in optical speckle potentials
where it was noted that the correlation length is several
times longer than the lattice spacing.

The paper is organized as follows. In Sec.\,2 we briefly outline
the Fermi-Hubbard model in a harmonic potential and disorder.
We then give the main ingredients of our density-functional
theory approach. Section 3 presents our results on site occupations
in the presence of correlated disorder. We conclude with a brief summary
in Sec.\,4. 

\section{Theory and Model}

We consider a two-component Fermi gas 
with $N$ atoms constrained to move 
under harmonic confinement of strength $V_2$ inside a 
disordered $1D$ optical lattice with unit lattice constant and $L$ 
lattice sites $i\in[1,L]$. The system is 
described by a single-band Hubbard Hamiltonian,
\begin{eqnarray}\label{eq:hubbard}
{\hat {\cal H}}&=&- t \sum_{i=1}^{L-1}\sum_\sigma 
({\hat c}^{\dagger}_{i\sigma}{\hat c}_{i+1\sigma}+{\rm H}.{\rm c}.)+
U\sum_{i=1}^L {\hat n}_{i\uparrow}{\hat n}_{i\downarrow}\nonumber\\
&+&V_2\sum_{i=1}^L (i-L/2)^2 {\hat n}_i
+\sum_{i=1}^L\varepsilon_i {\hat n}_i\,.
\end{eqnarray}
Here $t_{ij}=t>0$ if $i,j$ are nearest sites and zero 
otherwise, $\sigma=\uparrow,\downarrow$ is 
a pseudospin-$1/2$ label for two internal 
hyperfine states, 
${\hat n}_{i\sigma}= {\hat c}^{\dagger}_{i\sigma}{\hat c}_{i\sigma}$ 
is the pseudospin-resolved site occupation operator, and 
${\hat n}_i=\sum_\sigma {\hat n}_{i\sigma}$. The effect of disorder is 
simulated by the last term in Eq.~(\ref{eq:hubbard}). 
The above Hamiltonian without the confining potential and disorder 
term has been solved exactly using the Bethe Ansatz technique by 
Lieb and Wu.\cite{lieb_wu} The exact solution provides us with the ground-state energy 
at any coupling strength $u=U/t$ and filling $n=N/L$. 
In the noninteracting and unconfined
limit (i.e., without the $U$ and trap terms) one recovers the 
Anderson model\cite{anderson} studied intensively for the Anderson
localization problem.
The successful creation of optical lattices to study 
cold atomic systems has led to renewed interest in these
model systems, since the atomic systems offer a wide
range of control on the various parameters. 

In this work we generalize our previous study\cite{gao_short} on 
uncorrelated noise to Gaussian correlated disorder, defined as
\begin{equation}\label{eq:correlated}
\varepsilon_i=\varepsilon_i(\xi)=
\frac{1}{\sqrt{2\pi \xi}}\sum_{j=1}^{L}
\exp{\left[-\frac{(i-j)^2}{2\xi}\right]}{\cal W}_j~,
\end{equation}
where  $\xi$ is the correlation length (in units of the lattice spacing, 
which is set to unity throughout this work) and ${\cal W}_j$ is randomly 
chosen at each site $j$ from a uniform distribution in the range 
$[-W/2,W/2]$. Because of the following mathematical identity,

\begin{figure}[bt]
\begin{center}
\includegraphics[width=0.90\linewidth]{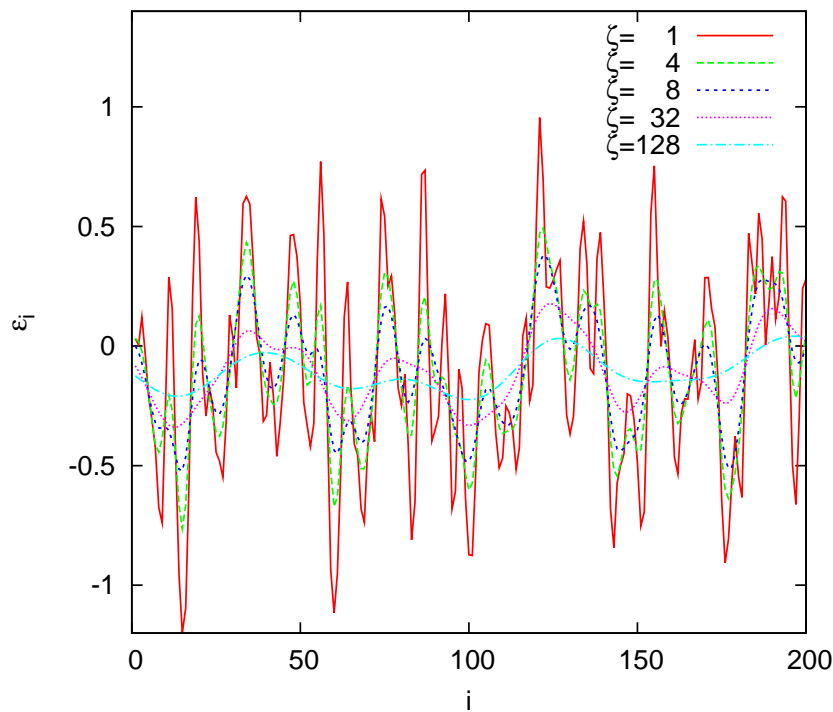}
\caption{(color online) The correlated disorder potential
$\varepsilon_i$ as a function of the site position for a lattice
of $L=200$ sites. The disorder strength is $W/t=3$. Different values
of $\xi$ are indicated in the legend. 
\label{fig:one}}
\end{center}
\end{figure}

\begin{equation}
\lim_{\xi\to 0}\frac{1}{\sqrt{2\pi \xi}}
\exp{\left[-\frac{(i-j)^2}{2\xi}\right]}=\delta_{ij}~,
\end{equation}
Eq.~(\ref{eq:correlated}) tends smoothly to uncorrelated 
({\it i.e.} white) noise in the limit $\xi \to 0$. We illustrate 
typical behavior of $\varepsilon_i$ over a lattice of $L=200$
sites in Fig.\,1 for some values of $\xi$. As the correlation length
increases the correlated disorder potential becomes smoother
with amplitude smaller than that in the uncorrelated case.

A particular set of values $\varepsilon_i$ is a realization of disorder.
Each realization defines an external potential ${\cal V}_i=V_2(i-N_s/2)^2
+\varepsilon_i$ which is the sum of a harmonic trap potential
and disorder.  the site-occupation functional 
theory\cite{soft} (SOFT)
to determine the site occupation $n_i=\langle\Psi |{\hat n}_i|\Psi\rangle$
where $\Psi$ is the ground-state of $\cal H$ for this particular 
disorder realization. The site occupation ${\cal N}_i$ is obtained 
by the disorder ensemble average, ${\cal N}_i=\langle\langle n_i\rangle
\rangle_{\rm dis}$. SOFT is the discrete or lattice version of 
the density functional theory which has been successfully applied
to the present system in the absence\cite{gao_long_2005} and 
presence\cite{gao_short} of uncorrelated disorder.
In the clean limit the local-density approximation which we adopt has 
been shown to be reliable through extensive comparisons with accurate 
quantum Monte Carlo calculations~\cite{gao_long_2005}. 
Local-density approximation based density-functional schemes 
for disordered systems have been employed to study 
the low-density compressibility anomaly in the two-dimensional
metal-insulator transition 
(see Ref.\,[\refcite{tanatar_ijmpb,droplet_state}]) 
and the statistical properties of $2D$ disordered quantum 
dots.\cite{hirose} 

The total energy is a unique functional of $n_i$ which can be written as
\begin{equation}
{\cal E}[n]=\sum_i \epsilon(n,u)+\sum_i {\cal V}_i(z_i)n(z_i)\, ,
\end{equation}
where $\epsilon(n,u)$ is the ground-state energy of the 
Hubbard Hamiltonian as obtained by Lieb and Wu.\cite{lieb_wu}
The Euler-Lagrange equation that follows from the above energy density is
\begin{equation}
\frac{\partial \epsilon}{\partial n}+v_{\rm KS}(z_i)={\rm constant}\, ,
\end{equation}
where the Kohn-Sham potential is
\begin{equation}
v_{\rm KS}=\frac{1}{2}Un+v_{\rm xc}(z_i)+{\cal V}_i(z_i)\, ,
\end{equation}
Within the local-density approximation, exchange-correlation potential 
is approximated by
\begin{equation}
v_{\rm xc}=\frac{\partial}{\partial n}\left[ \epsilon(n,u)-\epsilon(n,0)-
\frac{1}{4}Un^2\right]\, .
\end{equation}
The above set of equations allow us to obtain the site occupations $n_i$
and their disorder averages ${\cal N}_i$.

\section{Results and Discussion}

\begin{figure}
\begin{center}
\includegraphics[width=0.90\linewidth]{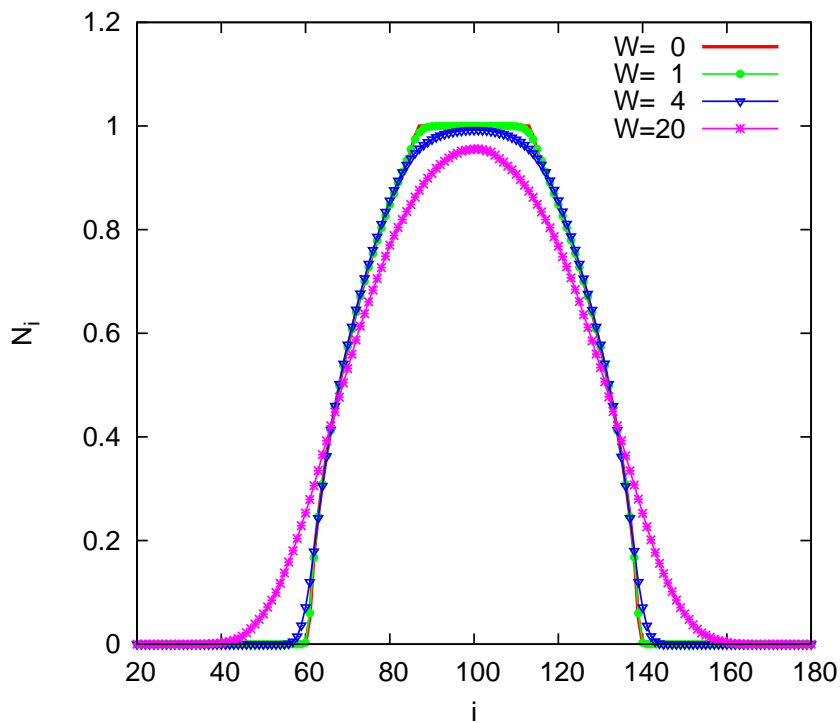}
\caption{(color online) Site occupation ${\cal N}_i$ as a function 
of site position $i$ for $N=60$ fermions with $u=4$, $\xi=4$, and 
$V_2/t=2.5\times 10^{-3}$ in a lattice with 
$L=200$ sites. The four curves have been calculated  
for different values of disorder strength: 
$W/t=0$ (solid line), $W/t=1$ (circles), $W/t=4$ (triangles), and 
$W/t=20$ (squares). 
\label{fig:two}}
\end{center}
\end{figure}

Before we describe the effects of disorder on the trapped lattice
fermions, we outline the various ground-state phases of a clean
system as obtained by previous numerical studies.\cite{gao_long_2005}
There are altogether five phases (${\cal A}\dots {\cal E}$) controlled
by the interaction strength $U/t$, number of fermions $N$, and
number of lattice sites $L$.
Phase $\cal A$ is a fluid with $0<n_i<2$. In phase $\cal B$ a
Mott insulated occupies the central region of the trap with $n_i=1$.
In phase $\cal C$ a fluid with $1<n_i<2$ is embedded in the Mott 
plateau. Phase $\cal D$ is a band insulator with $n_i=2$ surrounded
by fluid edges and embedded in the Mott plateau. Finally, in
phase $E$ a band insulator in the central region of the trap
coexists with fluid edges. The sketch of site occupations
in these phases was given in Ref.\,\refcite{gao_short}.

\begin{figure}
\begin{center}
\includegraphics[width=0.90\linewidth]{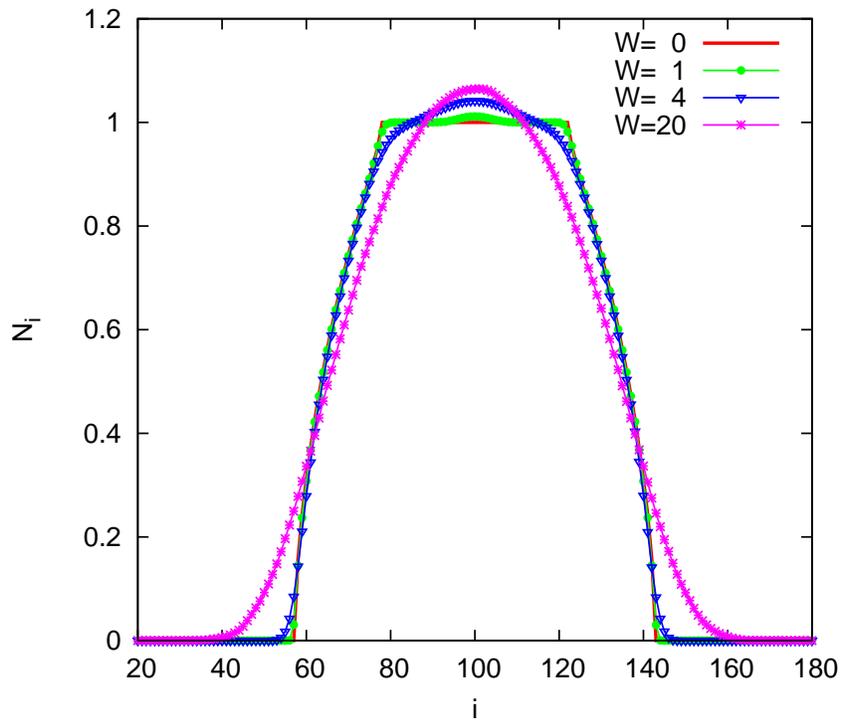}
\caption{(color online) Site occupation ${\cal N}_i$ as a function 
of site position $i$ for $N=70$ fermions with $u=4$, $\xi=4$, and 
$V_2/t=2.5\times 10^{-3}$ in a lattice with 
$L=200$ sites. The four curves have been calculated  
for different values of disorder strength: 
$W/t=0$ (solid line), $W/t=1$ (circles), $W/t=4$ (triangles), and 
$W/t=20$ (squares). 
\label{fig:three}}
\end{center}
\end{figure}

\begin{figure}
\begin{center}
\includegraphics[width=0.90\linewidth]{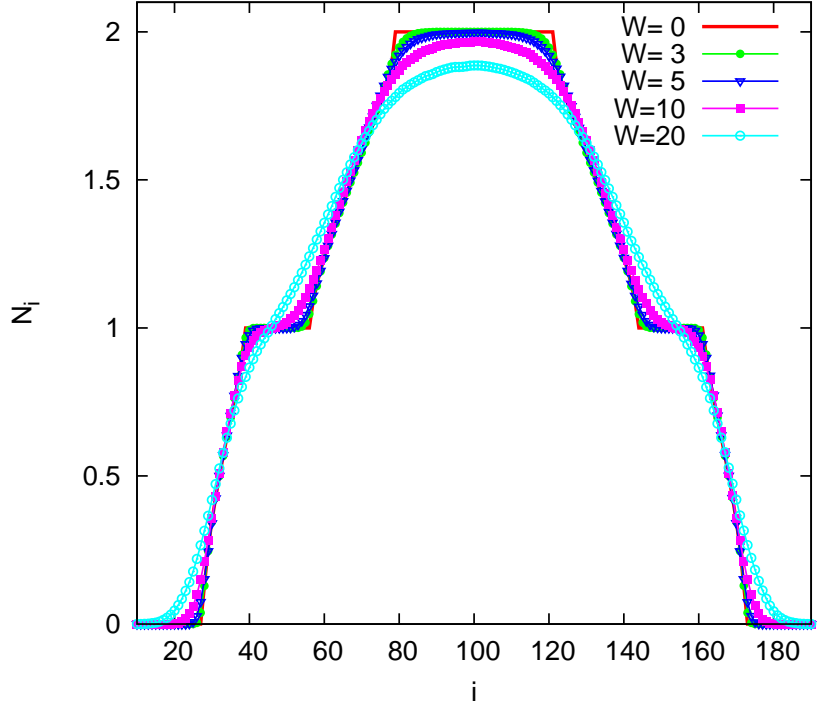}
\caption{(color online) Site occupation ${\cal N}_i$ as a function 
of site position $i$ for $N=200$ fermions with $u=8$, $\xi=4$, and 
$V_2/t=2.5\times 10^{-3}$ in a lattice with 
$L=200$ sites. The five curves have been calculated  
for different values of disorder strength: 
$W/t=0$ (solid line), $W/t=3$ (solid circles), $W/t=5$ (triangles), 
$W/t=10$ (squares), and $W/t=20$ (empty circles). 
\label{fig:four}}
\end{center}
\end{figure}

In Fig.\,2 we show the disorder-averaged site occupation ${\cal N}_i$ 
for a 
system of fermions with $N=60$ in an optical lattice with $L=200$
sites. The interaction strength is $u=4$ and the trap potential
is $V_2/t=1.5\times 10^{-3}$. Here we fixed the disorder correlation
length to be $\xi=4$ and varied the disorder strength $W/t$.
The clean system is in phase $\cal B$. As in the case of
uncorrelated disorder\cite{gao_short} we find that with
increasing $W/t$ the Mott insulating region is depleted and the
site occupation profile ${\cal N}_i$ broadens. However, the rate
of depletion and broadening are smaller than the uncorrelated disorder.
In other words, the Mott insulating region is more stable 
against the formation of a disordered fluid phase in the presence
of correlated disorder.

\begin{figure}
\begin{center}
\includegraphics[width=0.90\linewidth]{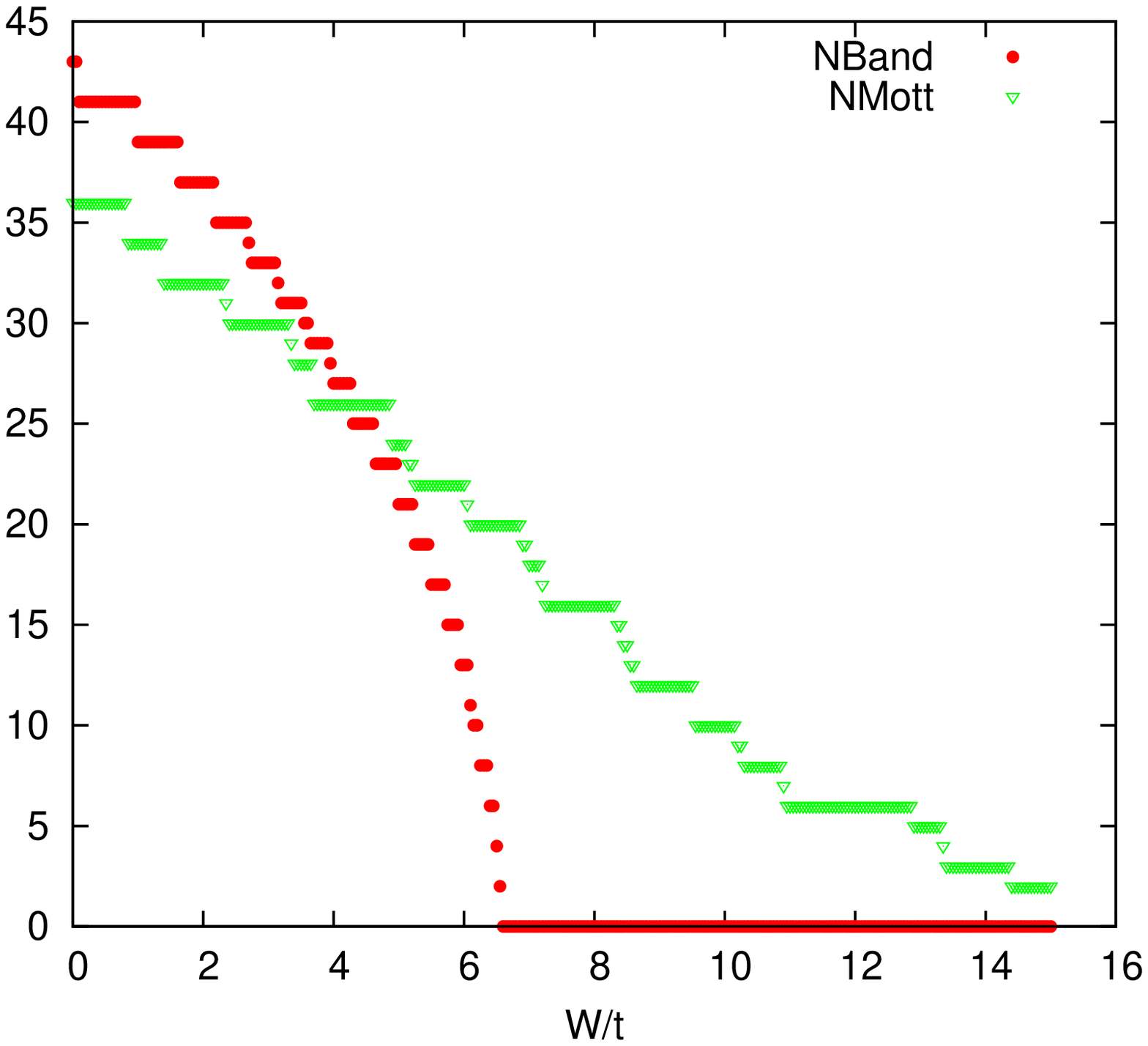}
\caption{(color online) The number of consecutive sites
$N_{\rm Mott}$ and $N_{\rm Band}$ such that $|{\cal N}_i-1|<0$
and $|{\cal N}_i-2|<0$, respectively, as a function of $W/t$. 
The parameters are the same as in Fig.\,4. 
\label{fig:five}}
\end{center}
\end{figure}

\begin{figure}
\begin{center}
\includegraphics[width=0.90\linewidth]{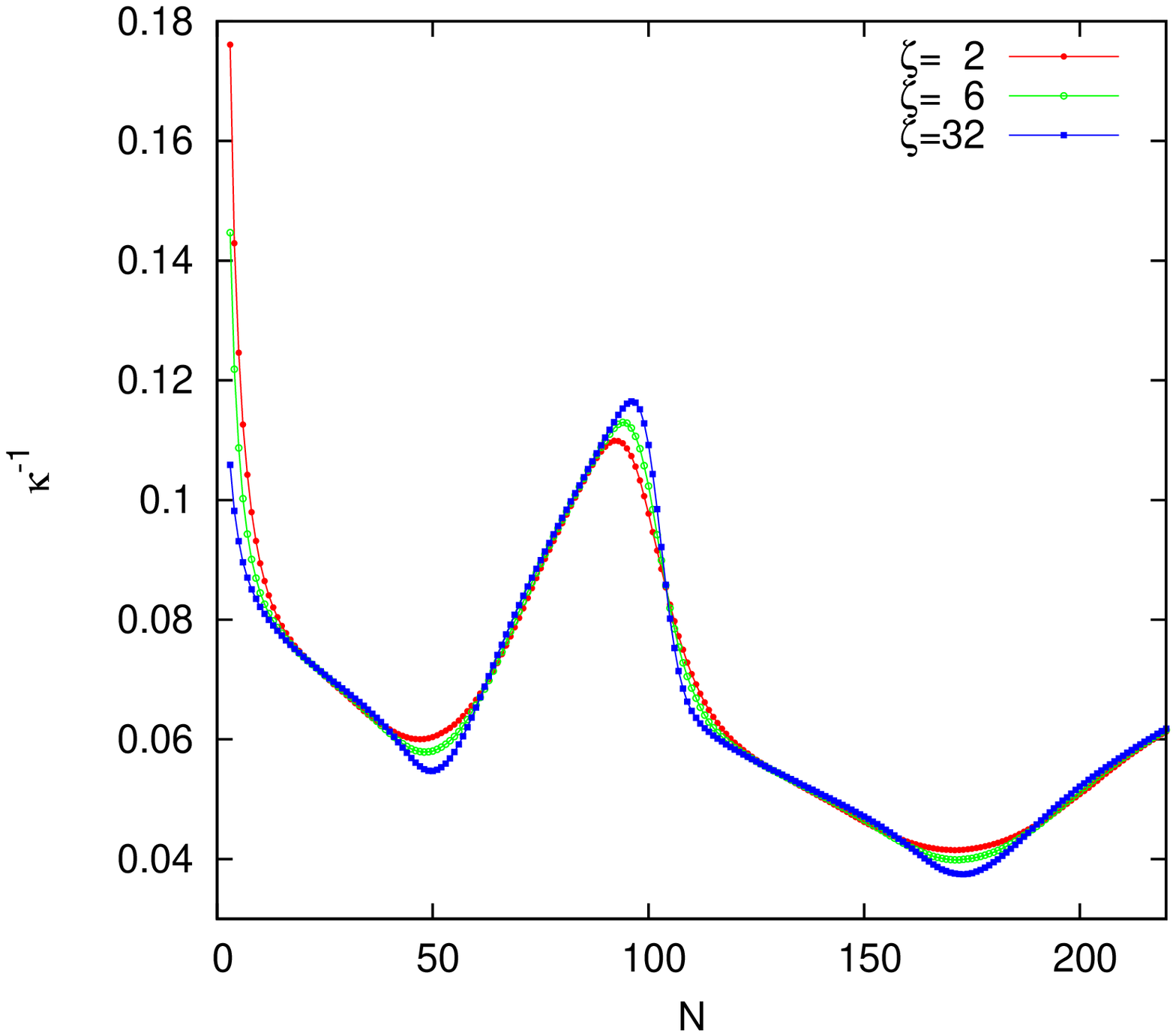}
\caption{(color online) The inverse compressibility $\kappa^{-1}$ 
(in units of $t$) in the presence of correlated disorder 
as a function of $N$ for $V_2/t=2.5\times 10^{-3}$, $W/t=5$, $u=8$, 
and $L=200$ lattice sites. 
The correlation length $\xi=0$ (circles), $\xi=6$ (squares), and
$\xi=32$ (triangles).
\label{fig:six}}
\end{center}
\end{figure}

In Fig.\,3 we display the disorder averaged site occupation ${\cal N}_i$
for a system with $N=70$ atoms keeping the rest of the parameters
the same as in Fig.\,2. $N=70$ is the critical number of atoms
at which the phase transition ${\cal B}\rightarrow {\cal C}$ occurs
in the clean limit. For weak uncorrelated disorder a fluid phase
with ${\cal N}_i>1$ is induced at the center of the trap. Essentially
the same behavior is observed for correlated disorder ($\xi=4$).
Sufficiently strong
$W/t$ destroys the Mott plateau completely and the system becomes
a disordered fluid with ${\cal N}_i<1$. We find that the critical
disorder strength for this to happen is much larger for correlated
disorder than the corresponding value of $W/t$ for uncorrelated
disorder.

In Fig.\,4 we show the site occupation for a strongly interacting
system ($U/t=8$) with correlated disorder ($\xi=4$). The disorder free
system is in the $\cal D$ phase discussed above. With increasing
$W/t$ the band insulating region at the center of the trap
is depleted. At the same time, the Mott insulating regions are
also slowly destroyed. To see more clearly the rate at which the 
band and Mott insulating regions are destroyed, we plot in
Fig.\,5 the number of consecutive sites $N_{\rm Mott}$ and
$N_{\rm Band}$ such that $|{\cal N}_i-1|<0$ and $|{\cal N}_i-2|<0$, 
respectively, as a function of $W/t$. Comparison with the same calculation
in the uncorrelated disorder case\cite{gao_short} reveals the
increased stability range of insulating regions for correlated
disorder.

The effect of disorder on 1D fermions and in particular transitions
between different phases can also be assessed
through the thermodynamic properties. For this purpose
we calculate the inverse compressibility defined as 
$\kappa^{-1}=\langle\langle N^2\delta\mu/\delta N\rangle\rangle_{\rm dis}$
where $\mu$ is the chemical potential.
In Fig.\,6 we show $\kappa^{-1}$ as a function of $N$
for various values of the disorder correlation length $\xi$.
The phase transitions can be identified as sharp kinks
in a clean system ($W/t=0$).  
Uncorrelated disorder has two main effects on $\kappa^{-1}$
as discussed in our previous work.\cite{gao_short}
Firstly, the sharp features indicating phase
transitions are smoothed out. Secondly, a large enhancement
of $\kappa^{-1}$ at low density is observed. This is reminiscent of
a similar behavior found in 2D electron systems. As $N$
decreases the atoms mostly occupy the deepest valleys in the 
disorder landscape, thus the high density regions in the system 
tend to become disconnected. For a given interaction strength
$u$ and low $N$ the system stiffens as disorder grows.
In the present situation we observe that with increasing 
correlation length $\xi$ the low density enhancement in
$\kappa^{-1}$ is diminished. The basic reason for this behavior
is that increasing $\xi$ makes the disorder potential 
more smooth and more shallow compared to the $\xi=0$ case.
Therefore the atoms are less localized in a correlated disorder
potential.

\section{Concluding remarks}

We have studied the one-dimensional Fermi-Hubbard model
in a harmonic confinement potential and in the presence of correlated
disorder. This is believed to represent cold fermionic atoms
in an optical lattice created by standing laser waves.
Our numerical calculations of the site occupations are based
on a lattice version of density functional theory in
which we make use of the exact solution of the one-dimensional
Hubbard model to treat the exchange-correlation effects.
Disorder affects the ground-state phases of the interacting
Fermi gases confined in a harmonic potential and an optical
lattice. The insulating regions appear to be stable against
both uncorrelated and correlated disorder. The anomalous enhancement of 
the stiffness observed at low density for uncorrelated disorder 
decreases with increasing correlation length.
We hope that our results on disorder effects will stimulate experimental
investigations with cold Fermi atoms in the future.

\section*{Acknowledgements}

G.\,X. was supported by NSF of China under Granr No. 10704066.
B.\,T. acknowledges the support by TUBITAK (No. 106T052),
TUBA and a travel grant from Julian Swchiwinger Foundation.


\section*{References}

\begin{thebibliography}{0}

\bibitem{MIT}
See for instance, E. Abrahams, S.V. Kravchenko, and M.P. Sarachik, 
Rev. Mod. Phys. {\bf 73}, 251 (2001);
B.L. Altshuler, D.L. Maslov, and V.M. Pudalov, 
Physica E {\bf 9}, 209 (2001).

\bibitem{helium_four}
J.D. Reppy, J. Low Temp. Phys. {\bf 87}, 205 (1992);
B. F{\aa}k, O. Plantevin, H.R. Glyde, and N. Mulders, Phys. Rev. Lett. 
{\bf 85}, 3886 (2000), and references therein.

\bibitem{fazio_2001}
R. Fazio and H. van der Zant, Phys. Rep. {\bf 355}, 235 (2001).	

\bibitem{spin_susceptibility}
A. Punnoose and A.M. Finkel'stein, Science {\bf 310}, 289 (2005);
A.A. Shashkin, S. Anissimova, M.R. Sakr, S.V. Kravchenko, 
V.T. Dolgopolov, and T.M. Klapwijk, 
Phys. Rev. Lett. {\bf 96}, 036403 (2006).	

\bibitem{compressibility_anomaly}
J.P. Eisenstein, L.N. Pfeiffer, and K.W. West, 
Phys. Rev. Lett. {\bf 68}, 674 (1992);
S. Ilani, A. Yacoby, D. Mahalu, and H. Shtrikman, 
{\it ibid}. {\bf 84}, 3133 (2000);
S.C. Dultz and H.W. Jiang, {\it ibid}. {\bf 84}, 4689 (2000).

\bibitem{tanatar_ijmpb} 
B. Tanatar, A.L. Suba{\c s}{\i}, K. Esfarjani,
and S.M. Fazeli, Int. J. Mod. Phys. B {\bf 21}, 2134 (2007).

\bibitem{speckle}
J.E. Lye, L. Fallani, M. Modugno, D.S. Wiersma, C. Fort, 
and M. Inguscio, Phys. Rev. Lett. {\bf 95}, 070401 (2005);
D. Cl\'ement, A.F. Var\'on, M. Hugbart, J.A. Retter, P. Bouyer, 
L. Sanchez-Palencia, D.M. Gangardt, G.V. Shlyapnikov, 
and A. Aspect, {\it ibid.} {\bf 95}, 170409 (2005);
C. Fort, L. Fallani, V. Guarrera, J.E. Lye, M. Modugno, 
D.S. Wiersma, and M. Inguscio, 
{\it ibid.} {\bf 95}, 170410 (2005);
for a review see V. Ahufinger, L. Sanchez-Palencia, A. Kantian, 
A. Sanpera, and M. Lewenstein, 
Phys. Rev. A {\bf 72}, 063616 (2005).

\bibitem{optical_lattices}
J.I. Cirac and P. Zoller, Science {\bf 301}, 176 (2003).

\bibitem{clement}
D. Clement, P. Bouyer, A. Aspect, and L. Sanchez-Palencia,
Phys. Rev. A: 77, 033631 (2008);
Y.P. Chen {\it et al}., e-print arXiv:0710.5187, 
to be published in Phys. Rev. A.

\bibitem{greiner}
M. Greiner, O. Mandel, T. Esslinger, T.W. H{\"a}nsch, and
I. Bloch, Nature {\bf 415}, 39 (2002).

\bibitem{hannover}
T. Schulte, S. Drenkelforth, J. Kruse, W. Ertmer, J. Arlt, 
K. Sacha, J. Zakrzewski, and M. Lewenstein, 
Phys. Rev. Lett. {\bf 95}, 170411 (2005).

\bibitem{moritz_2005}
H. Moritz, T. St\"oferle, K. G\"unter, M. K\"ohl, and T. Esslinger, 
Phys. Rev. Lett. {\bf 94}, 210401 (2005).

\bibitem{fermi_lattice}
W. Hofstetter, C.I. Cirac, P. Zoller, E. Demmler, and M.D. Lukin,
Phys. Rev. Lett. {\bf 89}, 220407 (2002);
S. Trebst, U. Schollw{\"o}k, M. Troyer, and P. Zoller, Phys. Rev. Lett.
{\bf 96}, 250402 (2006).

\bibitem{giorgini}
I. Bloch, J. Dalibard, and W. Zwerger, arXiv:0704.3011v2, to be published 
in Rev. Mod. Phys.;	
L. Giorgini, L.P. Pitaevskii, and S. Stringari, e-print arXiv:0706.3360v2,
to be published in Rev. Mod. Phys. 

\bibitem{soft}
K. Sch\"onhammer, O. Gunnarsson, and R.M. Noack, 
Phys. Rev. B {\bf 52}, 2504 (1995);
N.A. Lima, M.F. Silva, L.N. Oliveira, and K. Capelle, 
Phys. Rev. Lett. {\bf 90}, 146402 (2003).

\bibitem{lieb_wu}
E.H. Lieb and F.Y. Wu, Phys. Rev. Lett. {\bf 20}, 1445 (1968).

\bibitem{repulsive}
M. Rigol, A. Muramatsu, G.G. Batrouni, and R.T. Scalettar, 
Phys. Rev. Lett. {\bf 91}, 130403 (2003);
M. Rigol and A. Muramatsu, Phys. Rev. A {\bf 69}, 053612 (2004); 
Opt. Commun. {\bf 243}, 33 (2004);
X.-J. Liu, P.D. Drummond, and H. Hu, Phys. Rev. Lett. 
{\bf 94}, 136406 (2005); 
V.L. Campo, Jr. and K. Capelle, Phys. Rev. Lett. {\bf 72}, 061602 (2005).

\bibitem{gao_long_2005}
Gao Xianlong, M. Polini, M.P. Tosi, V.L. Campo, Jr., K. Capelle, 
and M. Rigol, Phys. Rev. B {\bf 73}, 165120 (2006).

\bibitem{gao_short} 
Gao Xianlong, M. Polini, B. Tanatar, and M.P. Tosi, Phys. Rev. B
{\bf 73}, 161103 (2006).

\bibitem{anderson}
P.W. Anderson, Phys. Rev. {\bf 109}, 1492 (1958).	

\bibitem{droplet_state}
J. Shi and X.C. Xie, Phys. Rev. Lett. {\bf 88}, 086401 (2002).	

\bibitem{hirose}
K. Hirose, F. Zhou, and N.S. Wingreen, 
Phys. Rev. B {\bf 63}, 075301 (2001);
K. Hirose and N.S. Wingreen, {\it ibid.} {\bf 65}, 193305 (2002);
H. Jiang, D. Ullmo, W. Yang, and H.U. Baranger, 
{\it ibid.} {\bf 69}, 235326 (2004);
E. R\"as\"anen and M. Aichinger, {\it ibid.} {\bf 72}, 045352 (2005).








\end{thebibliography}
\end{document}